RESEARCH ARTICLE

# Rule Extraction Based on Extreme Learning Machine and an Improved Ant-Miner Algorithm for Transient Stability Assessment

Yang Li*, Guoqing Li, Zhenhao Wang

School of Electrical Engineering, Northeast Dianli University, Jilin, Jilin, P.R.China

* liyang@nedu.edu.cn

## Abstract

In order to overcome the problems of poor understandability of the pattern recognition-based transient stability assessment (PRTSA) methods, a new rule extraction method based on extreme learning machine (ELM) and an improved Ant-miner (IAM) algorithm is presented in this paper. First, the basic principles of ELM and Ant-miner algorithm are respectively introduced. Then, based on the selected optimal feature subset, an example sample set is generated by the trained ELM-based PRTSA model. And finally, a set of classification rules are obtained by IAM algorithm to replace the original ELM network. The novelty of this proposal is that transient stability rules are extracted from an example sample set generated by the trained ELM-based transient stability assessment model by using IAM algorithm. The effectiveness of the proposed method is shown by the application results on the New England 39-bus power system and a practical power system — the southern power system of Hebei province.

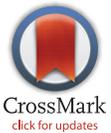

🔓 OPEN ACCESS

**Data Availability Statement:** All relevant data are within the paper.

**Funding:** The authors received no specific funding for this work.

**Competing Interests:** The authors have declared that no competing interests exist.

## Introduction

Transient stability is concerned with the stability of the power system to maintain synchronism when subjected to a severe disturbance, such as a short circuit on a transmission line [1]. Transient stability assessment (TSA) has been recognized as an important issue to ensure the secure operation of power systems. With interconnection of large-scale power grids, electricity market reform and growing presence of large-scale intermittent renewable energy, the dynamic behaviors of power systems are becoming more complex and difficult to be controlled, with more serious consequences resulted from transient instability [2]. The available TSA methods, such as time domain simulation methods [3], energy function based methods [4] and the extended equal-area criterion [5], can not meet the demands of online applications required by the modern power systems. With the rapid development of computational intelligence, the pattern recognition-based TSA (PRTSA) methods have shown much potential for on-line application to power systems [6–9]. Its main task is to establish a mapping relationship between system state variables and system stability conclusions. Compared to the other TSA methods, the PRTSA methods have a lot of advantages, such as strong learning ability, fast assessment speed and acquisition of potentially useful information. They have a good prospect in the field of the on-line security and stability analysis of power systems.

However, it has also been observed that the current researches of PRTSA mainly focus on the application of machine learning technique such as artificial neural network (ANN) and support vector machine (SVM) to make stability classification of power systems [8, 9]. Although the above PRTSA methods can perform well for TSA, the black-box nature of the generated classifiers makes them rather incomprehensible and opaque to humans [10, 11], since the predictive models are described as complex mathematical functions. This opacity property is not conducive to understand and analysis the stability problem of power systems from the mechanism [12], and prevents them from being used in practical applications.

Rule extraction is an effective way to make the "black-box" PRTSA approaches have incomprehensibility, whose purpose is to express the knowledge that is implicit in the learning machine in an easily understandable way. Unfortunately, there has been a little research on this issue. In [10, 11], decision tree is employed to extract rules, but the assessment results are sensitive to the construction of samples with the poor generalization ability and robustness as well. IF–THEN rules are extracted from the trained multilayer perceptron (MLP) by scrutinising the weights between hidden-output layer and the weights between input-hidden layer in [12], however the resulting rules are not fine enough.

Extreme learning machine (ELM) proposed by Huang [13, 14] is a new learning scheme for single hidden layer feed forward networks (SLFNs). Compared to other traditional machine learning techniques such ANN and SVM, ELM has better generalization performance with a much faster learning speed, which has been widely used in a lot of engineering applications [15, 16].

However, the acquired knowledge from ELM is contained in the connection weights, and the reasoning process can not be given a clear explanation, which limits further application of ELM in data mining and engineering. It's contribution to enhance the understandability and interpretation of the reasoning process by representing the knowledge contained in ELM in the form of rules, but so far none of literature on this issue has been reported.

Ant-miner algorithm is a new rule mining algorithm [17] with good robustness and ability to find optimal solutions, which has been successfully applied in engineering applications [18]. Meanwhile, wide-area measurement systems (WAMS) make it possible to obtain the synchronized real-time state information, and this brings new ideas and opportunities to transient stability assessment and prediction [19–22].

A novel rule extraction method based on ELM and improved Ant-miner (IAM) algorithm is proposed in this paper. The novelty of this proposal is that transient stability rules are extracted from an example sample set generated by the trained ELM-based TSA model by using IAM algorithm. Based on a view of function replacement, if the generated example samples reflecting the response characteristic of the trained model are sufficient enough and cover the entire sample space with the uniform distribution, the obtained rules will have the similar functions with the original ELM. In this way, the presented approach improves the understandability and interpretability of the "black-box" PRTSA models.

The remainder of this paper is structured as follows. First, the basic principles of ELM and Ant-miner algorithm are respectively introduced in brief. Details of the proposed rule extraction scheme using ELM and IAM for TSA are presented next. Application of the proposed scheme is demonstrated using the New England 39-bus power system and a practical power system—the southern power system of Hebei province, and finally the conclusions are made.

## Principle of ELM and Ant-miner Algorithm

Principle of ELM

For $N$ arbitrary distinct samples $(\mathbf{x}_i, \mathbf{y}_i) \in \mathbf{R}^n \times \mathbf{R}^m$, where $\mathbf{x}_i = [x_{i1}, x_{i2}, \cdots, x_{in}]^\mathrm{T} \in \mathbf{R}^n$ is the feature vector and $\mathbf{y}_i = [y_{i1}, y_{i2}, \cdots, y_{im}]^\mathrm{T} \in \mathbf{R}^m$ is the target vector, standard SLFNs with $L$ hidden nodes can be mathematically modeled as

$$\sum_{i=1}^{L} \beta_i G(\mathbf{w}_i \cdot \mathbf{x}_j + d_i) = \mathbf{y}_j, j = 1, \cdots, N. \tag{1}$$

where $\mathbf{w}_i = [w_{i1}, w_{i2}, \cdots, w_{in}]^\mathrm{T}$ is the weight vector connecting the $i$th hidden node and the input nodes, $\beta_i = [\beta_{i1}, \beta_{i2}, \cdots, \beta_{im}]^\mathrm{T}$ is the weight vector connecting the $i$th hidden node and the output nodes, and $d_i$ is the threshold of the $i$th hidden node, $G(\cdot)$ is the activation function.

Eq (1) can be written compactly as

$$\mathbf{H}\boldsymbol{\beta} = \mathbf{Y} \tag{2}$$

where $\mathbf{H}$ is the hidden layer output matrix of the neural network,

$$\mathbf{H}(\mathbf{w}_1, \cdots, \mathbf{w}_L, d_1, \cdots, d_L, \mathbf{x}_1, \cdots, \mathbf{x}_N) = \begin{bmatrix} G(\mathbf{w}_1 \cdot \mathbf{x}_1 + d_1) & \cdots & G(\mathbf{w}_L \cdot \mathbf{x}_1 + d_L) \\ \vdots & \cdots & \vdots \\ G(\mathbf{w}_1 \cdot \mathbf{x}_N + d_1) & \cdots & G(\mathbf{w}_L \cdot \mathbf{x}_N + d_L) \end{bmatrix}_{N \times L}, \text{the}$$

unique parameter needed to be tuned is $\beta = [\beta_1, \cdots, \beta_L]^\mathrm{T}$, $\mathbf{Y} = [\mathbf{y}_1, \cdots, \mathbf{y}_L]^\mathrm{T}$.

ELM is to minimize the training error as well as the norm of the output weights [13, 14]

$$\text{Minimize} : \|\mathbf{H}\boldsymbol{\beta} - \mathbf{Y}\|^2 \text{ and } \|\boldsymbol{\beta}\| \tag{3}$$

The minimal norm least square solution of (2) is as follows.

$$\hat{\boldsymbol{\beta}} = \mathbf{H}^\dagger \mathbf{Y}. \tag{4}$$

where $\mathbf{H}^\dagger$ is the Moore-Penrose generalized inverse of matrix $\mathbf{H}$.

Given a training set, the activation function and the hidden nodes, learning process of ELM is as follows: (a) randomly generated parameters of the hidden layer nodes $(\mathbf{w}_i, d_i)$, $i = 1, \cdots, L$; (b) calculate the hidden layer output matrix $\mathbf{H}$; (c) calculate the output weights $\boldsymbol{\beta}$.

### Introduction of Ant-miner algorithm

Ant-miner algorithm is to simulate the process of extraction rules into the process of ants foraging, and the optimal path is chosen as the optimal classification rules [17]. The specific steps are described as follows.

Step 1: Initialization of pheromones. The initial path of the pheromone $\tau_{ij}$ is set as follows.

$$\tau_{ij}(t = 0) = \frac{1}{\sum_{i=1}^{a} b_i} \tag{5}$$

where $t$ and $a$ are respectively the number of iterations and attributes, and $b_i$ is the number of values in the domain of the $i$th attribute. Pheromone level is updated in two phases: evaporation and reinforcement. Evaporation is accomplished by a pheromone evaporation rate $\rho$, and reinforcement of the pheromone trail is only applied to the best ant's path.

Step 2: Selection of attributes. Let $term_{ij}$ be a rule condition of the form $A_i = V_{ij}$, where $A_i$ is the $i$th attribute value and $V_{ij}$ is the $j$th value of the domain of $A_i$. The probability that $term_{ij}$ is selected to be added to the current partial rule is determined by the decision $P_{ij}$:

$$P_{ij}(t) = \frac{\tau_{ij}(t) \times \eta_{ij}}{\sum_{i=1}^{a} \sum_{j=1}^{b_i} (\tau_{ij}(t) \times \eta_{ij})} \tag{6}$$

where $\tau_{ij}(t)$ is the pheromone of $term_{ij}$ in the $t$th iteration, the heuristic function $\eta_{ij}$ is represented by (7), (8).

$$\eta_{ij} = \frac{\log_2(k) - InfoT_{ij}}{\sum_{i=1}^{a} \sum_{j=1}^{b_i} (\log_2(k) - InfoT_{ij})} \tag{7}$$

$$InfoT_{ij} = -\sum_{e=1}^{k} \left[ \frac{freqT_{ij}^e}{|T_{ij}|} \right] \times \log_2 \left[ \frac{freqT_{ij}^e}{|T_{ij}|} \right] \tag{8}$$

where $k$ is the number of categories, $InfoT_{ij}$ is the information entropy of $term_{ij}$, $|T_{ij}|$ is the samples number of the partition $T_{ij}$, $freqT_{ij}^e$ is the samples number of class $e$ in $T_{ij}$.

Step 3: Rule pruning. Aiming at the problem of over-fitting, the obtained rules are needed to be pruned. The effectiveness of the rules $Q$ is determined by (9).

$$Q = \frac{TP}{TP + FN} \times \frac{TN}{TN + FP} \tag{9}$$

where, $TP$ (true positives) is the number of cases covered by the rule that have the class predicted by the rule, $FP$ (false positives) is the number of cases covered by the rule that have a class different from the class predicted by the rule, $FN$ is (false negatives) the number of cases that are not covered by the rule but that have the class predicted by the rule, and $TN$ (true negatives) is the number of cases that are not covered by the rule and that do not have the class predicted by the rule.

Step 4: Pheromone updating. Pheromone is updated according to (10).

$$\tau_{ij}(t+1) = (1 - \rho)\tau_{ij}(t) + \left( \frac{Q}{1+Q} \right) \times \tau_{ij}(t) \tag{10}$$

where $\rho$ is the pheromone evaporation rate.

Step 5: Choose the best rule $R_{best}$, and adding it to the rule sets.

Step 6: Delete the training samples covered by the existing rules.

Step 7: Repeat step 1~6, until the number of training samples is not bigger than the maximum number of the preset uncovered samples.

## Rule Extractions for TSA

### Steps of rule extraction

From a functional point of view, a rule extraction method based on ELM and IAM is proposed for TSA as shown in Fig 1. The proposed method focuses on the mapping relationship between the state information and the stability result of power systems, and emphasizes the ability to reproduce the function of the ELM classifier, which does not consider the type and structure of ELM. The basic idea of the proposed method is to regard the trained ELM as a new sample space, and then use IAM algorithm to extract the hidden knowledge into rules with good understandability. In this way, the classification rules are generated to functionally replace the original ELM network.

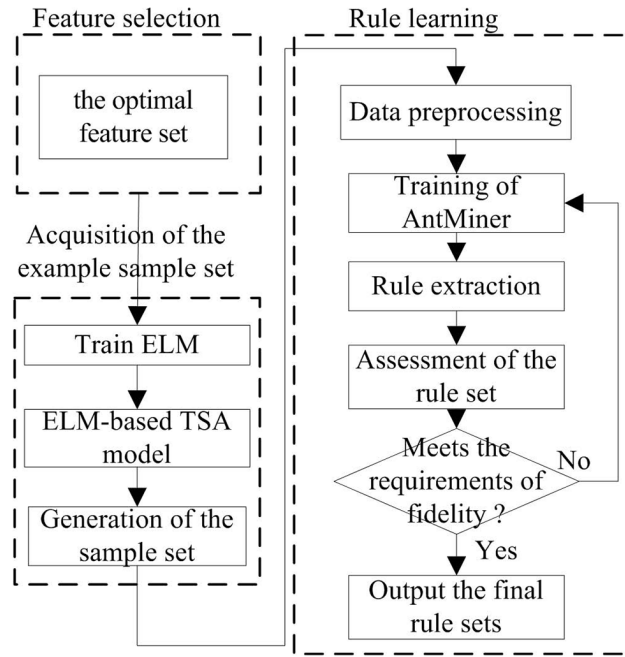

**Fig 1. Flowchart of ELM-based rule extraction.**

The proposed rule extraction approach can be divided into three steps, comprising feature selection, acquisition of the example sample set and rule learning.

## Feature selection

As is well known, feature selection is an issue of paramount importance for PRTSA. In [23], a feature selection method based on kernelized fuzzy rough sets (KFRS) and the memetic algorithm is proposed for TSA. By defining a KFRS-based generalized classification function as the separability criterion, the memetic algorithm based on binary differential evolution and Tabu search is employed to obtain the optimal feature subsets with the maximized classification capability (see [23] for further details).

The approach presented here uses the same feature selection method in [23], and the optimal feature subsets obtained for the New England 39-bus system and the southern power system of Hebei province are used as the input features in this study, as respectively shown in Tables 1 and 2. Here, $t_0$ and $t_{cl}$ denote the fault occurrence and clearing time in turn, $t_{cl+3c}$, $t_{cl+6c}$ and $t_{cl+9c}$ respectively denotes the 3-rd, 6th and 9-th cycle after the fault clearance.

## Acquisition of the example sample set

An example consists of the input mode and output mode. If an example is judged by using the trained ELM model, the assessment result is used as the output mode. Then, a new example can be obtained by composing the obtained output mode and the original input mode, which reflects the response characteristic of ELM in a certain extent [24]. If the examples are sufficient enough and cover the entire sample space, the rules obtained will have the similar functions

**Table 1. The input features for the New England 39-bus test system.**

| No. | Input features |
| --- | --- |
| **Tz1** | Mean value of all the mechanical power before the fault incipient time |
| **Tz2** | Mean value of all the initial acceleration power |
| **Tz3** | Rotor angular velocity of the machine with the biggest difference relative to the centre of inertia at $t_{cl+3c}$ |
| **Tz4** | Rotor angle of the machine with the biggest difference relative to the centre of inertia at $t_{cl+6c}$ |
| **Tz5** | Rotor angular velocity of the machine with the biggest difference relative to the centre of inertia at $t_{cl+6c}$ |
| **Tz6** | Rotor angle of the machine with the biggest difference relative to the centre of inertia at $t_{cl+9c}$ |
| **Tz7** | Rotor angular velocity of the machine with the biggest difference relative to the centre of inertia at $t_{cl+9c}$ |

doi:10.1371/journal.pone.0130814.t001

with the original ELM, i.e., these rules can describe the functions of the original network. The steps of generating the example sample set are listed as follows.

Step 1: Based on the training set A with class labels, the TSA model is obtained by using ELM.

Step 2: Determine the range of the input features, and then generate a random data set $B_1$ (the input modes) without the corresponding class labels in the range. In addition, one should to note that the data set $B_1$ has the same input features with the training set A, and their values are different.

Step 3: The obtained data set $B_1$ is assessed by using the trained ELM-based TSA model, and then the corresponding class labels (the output modes) are obtained.

Step 4: Finally, the example sample set B used for the follow-up rule learning can be acquired by combining with the input and output modes, which are respectively obtained in Step 2 and Step 3.

## Rule learning

**IAM algorithm.** "No free lunch" theorem [25] shows that there is no optimization algorithm is optimal for any problems such as global optimization ability and convergence speed. The original Ant-Miner algorithm is improved from two aspects in this paper. On the one

**Table 2. The input features for the southern power system of Hebei province.**

| No. | Input features |
| --- | --- |
| **Tz1** | Mean value of all the initial acceleration power |
| **Tz2** | Maximum value of all the rotor kinetic energies at $t_{cl}$ |
| **Tz3** | Rotor angle of the machine with the biggest difference relative to the centre of inertia at $t_{cl+3c}$ |
| **Tz4** | Maximum value of the difference of rotor angles at $t_{cl+3c}$ |
| **Tz5** | Kinetic energy of the machine with the maximum rotor angle at $t_{cl+3c}$ |
| **Tz6** | Rotor angle of the machine with the biggest difference relative to the centre of inertia at $t_{cl+6c}$ |
| **Tz7** | Maximum value of the difference of rotor angles at $t_{cl+6c}$ |
| **Tz8** | Rotor angular velocity of the machine with the biggest difference relative to the centre of inertia at $t_{cl+6c}$ |
| **Tz9** | Rotor angle of the machine with the biggest difference relative to the centre of inertia at $t_{cl+9c}$ |
| **Tz10** | Maximum value of the difference of rotor angles at $t_{cl+9c}$ |
| **Tz11** | Rotor angular velocity of the machine with the biggest difference relative to the centre of inertia at $t_{cl+9c}$ |

doi:10.1371/journal.pone.0130814.t002

hand, an adaptive tuning strategy is adopted to tune the pheromone evaporation rate $\rho$; on the other hand, the heuristic function is improved to reduce the computational overhead.

**(1) Improvement of pheromone evaporation rate** $\rho$. In Ant-miner algorithm, the control parameter $\rho$ plays an important role in the performance of Ant-miner algorithm, so an adaptive tuning strategy is adopted here. If $\rho$ is bigger, the algorithm is not easy to fall into local optimum, but its convergence speed is slow; otherwise, its convergence speed is fast, but it is easy to fall into local optimum. In this paper, a self-adaptive adjustment control way is employed to improve the performance of the algorithm,

$$1 - \rho(t+1) = \begin{cases} 0.95 \times (1 - \rho(t)), & \text{if } 0.95 \times (1 - \rho(t)) \geq \rho_{\min} \\ \rho_{\min}, & \text{else} \end{cases} \tag{11}$$

where $\rho_{\min}$ is the minimum of $\rho$.

By setting the dynamic parameter $\rho$, the useful information in the last search can be preserved, which facilitates a more finer search in a better area. By this means, not only the convergent speed but also the searching ability are enhanced.

**(2) Improvement of the heuristic function.** The heuristic value in AntMiner is defined as an information theoretic measure in terms of the entropy. Unfortunately, the information entropy-based heuristic function is complex and time-consuming. With the assumption that the small induced errors are compensated by the pheromone level [26], a density-based heuristic function is employed as shown in (12), which makes IAM computationally less expensive without a significant degradation of the stated performance.

$$\eta_{ij}(t) = \frac{|T_{ij}(t)|}{|Ts(t)|} \tag{12}$$

where $|T_{ij}|$ and $|Ts|$ are the first iteration $t$, respectively, the total number of samples in the sample number and the division $T_{ij}$ of the training set.

**Process of IAM algorithm.** An iteration of the IAM Algorithm mainly consists of three steps, comprising rule construction, rule pruning, and pheromone updating, detailed as shown in Fig 2.

**Rule learning based on IAM.** The rule learning steps are listed as follows:

Step 1: Data pre-processing. The z-score standardization method is used as the data pre-processing method for the obtained sample set.

$$f' = (f - \bar{F})/\sigma_F \tag{13}$$

where $\bar{F}$ and $\sigma_F$ are the mean and standard deviation of any feature $F$ in sample data, respectively; $f'$ is the normalized value of $f, f \in F$.

Step 2: Based on the trained ELM-based TSA model, an example sample set $S$ can be obtained. To approximate the functions of the original ELM network, the obtained samples should be sufficient enough and cover the entire sample space with the uniform distribution.

Step 3: If there exist some examples with a same attributes combination belong to the same class in $S$, then a new rule $R_u$ is established, i.e. the attributes combination and the class are respectively used for the rule antecedent and rule consequent.

Step 4: Based on the trained IAM-based TSA model, a new example sample set $S^*$ is generated to justify the fidelity of the rule set. If the fidelity of the rule set meets the requirements, then the rule $R_u$ is retained in the discovered rule list, and the covered cases by the rule $R_u$ is removed from $S$; otherwise, $R_u$ is eliminated from the rule list.

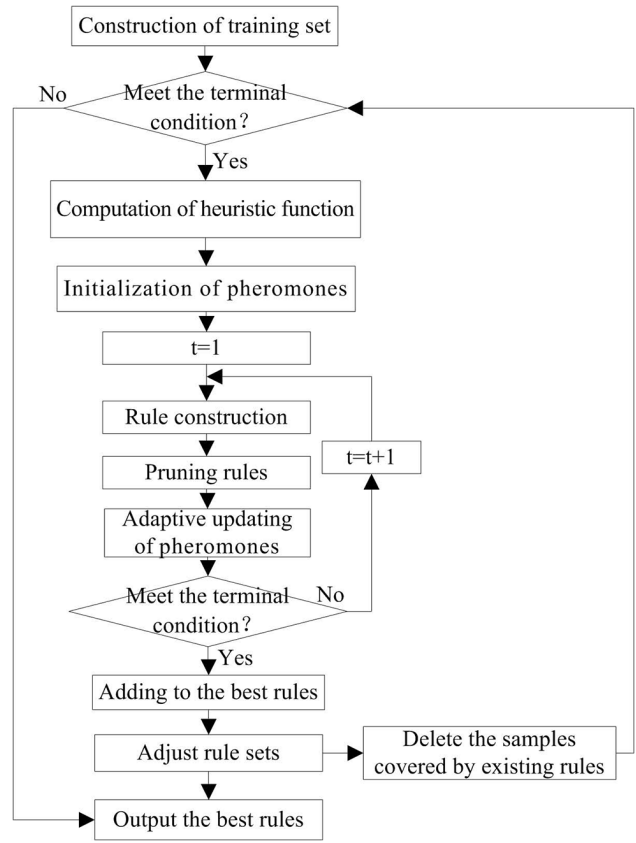

**Fig 2. Flowchart of IAM algorithm.**

Step 5: If $S = \emptyset$, then stop the learning process; otherwise, go to Step 3 and repeat the process. This makes the generated rules gradually approaching the functions of the original ELM classifier.

Step 6: Output the generated rule set.

## Case Studies

The effectiveness of the proposed method is tested on the New England 39-bus power system and a practical power system—the southern power system of Hebei province. All programs are implemented in MATLAB on a PC platform with the master frequency 1.81 GHz and main memory 1 GB.

In order to properly assess the performance of the proposed predictive model, the well-known model validation technique, cross-validation, is employed in the following case studies, as it provides a nearly unbiased estimate of the generalization ability of predictive models by avoiding overfitting and underfitting.

## Case 1—The New England 39-bus power system

The New England 39-bus power system is the widely-used test system for TSA studies reported in the literature [10–12, 22]. The one-line diagram of the power system is shown in Fig 3.

**Generation of the sample sets.** Extensive time domain simulation work has been carried out to create the training and test sample sets. The simulation is done with the four-order

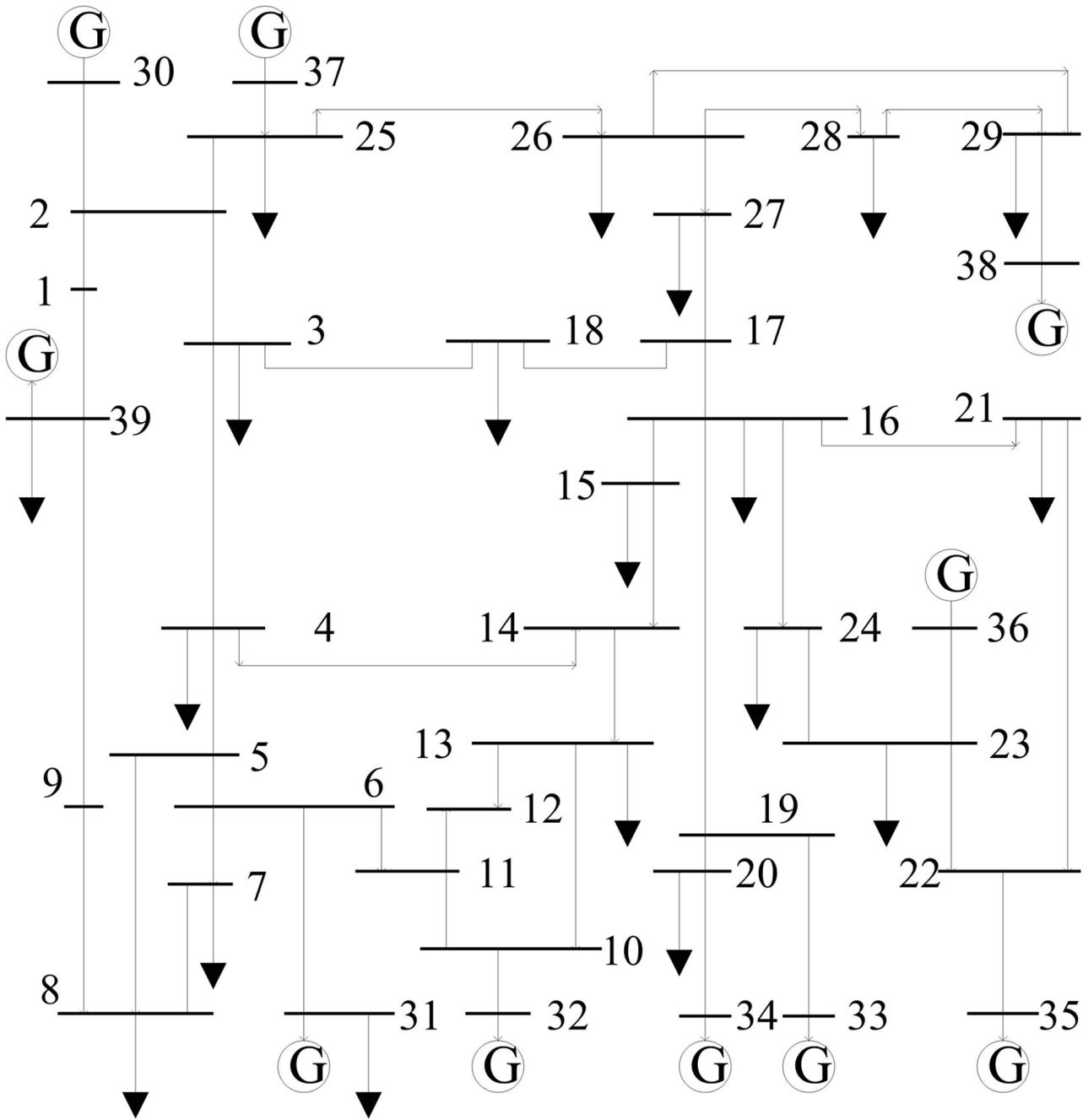

**Fig 3. New England 39-bus test system.**

doi:10.1371/journal.pone.0130814.g003

machine model and the IEEE DC1 excitation system model, as well as the constant impedance load model. A three-phase short-circuit faults is created at instant 0.1 s and cleared at 0.2 s. A successful reclosure of the faulted line is applied after fault clearance with no topology change from the fault. A total of 4800 arbitrary samples at 80 different fault locations are created under 75%, 80%, 85%, . . . . . ., 130% of the basic load levels. Corresponding to each loading level, 5 different generator outputs are randomly set.

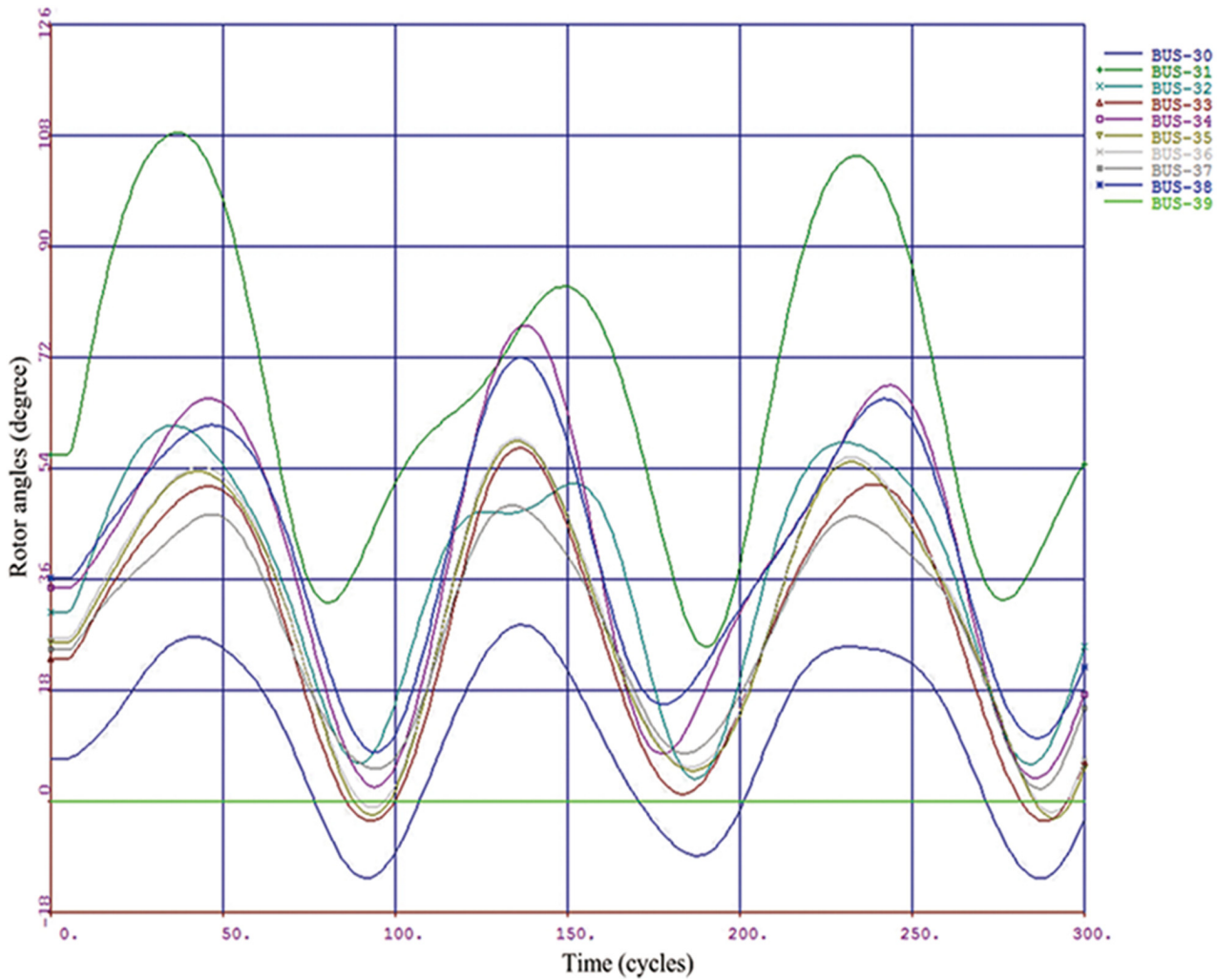

**Fig 4. Transient stable case.**

A class label "-1" or "+1" is assigned to each sample according to maximum relative rotor angle deviation during the transient period. If the maximum relative rotor angle deviation exceeded 360 degree [22, 23], the class label is given as "-1" and the system is considered to be transiently unstable; otherwise, the label is given as "+1" and the system is stable. In Figs 4 and 5, a transient stable case and an unstable case are respectively shown by plotting the rotor angle swing curves.

**Parameter settings.** In the proposed approach, there are only two parameters, the evaporation rate $\rho$ and the number of ants, needed to be set. To obtain the optimal parameters, large amounts of experiments over a range of parameter settings have been carried out with the results shown in Figs 6–8. Fig 6 shows the influence of these parameters on accuracy with a wider line used for the experiments with 400 ants and varying $\rho$, and the experiments with $\rho$ set to 0.85 and varying number of ants. Fig 7 and Fig 8 show the surface lines for a constant evaporation rate (0.85) and constant number of ants (400), respectively.

It might be taken for granted that the better results will be obtained if the parameters are higher, since more ants mean that more candidate rules are generated and increasing $\rho$ will cause the convergence process to be slower. However, from a certain threshold on, an

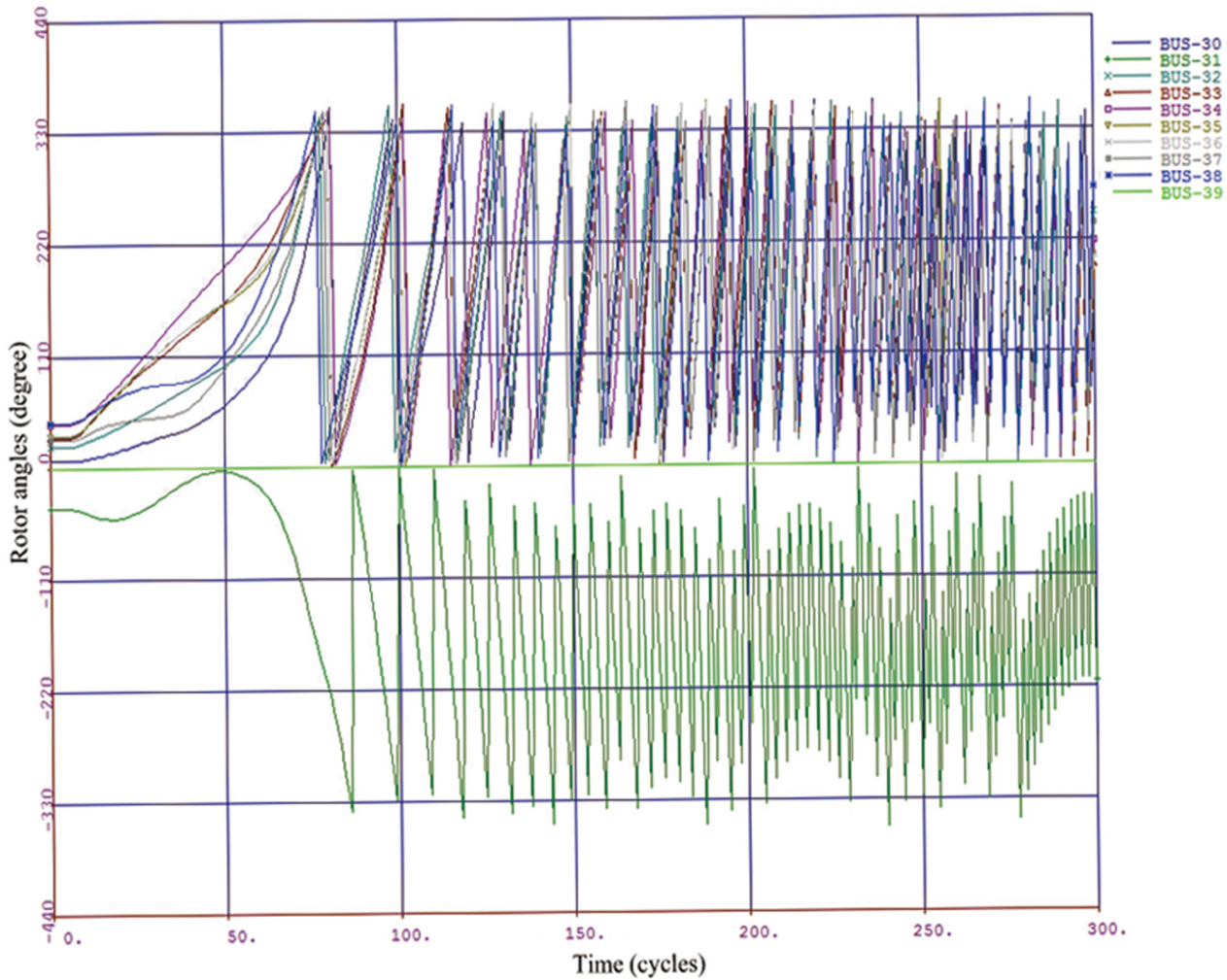

**Fig 5. Transient unstable case.**

doi:10.1371/journal.pone.0130814.g005

interesting phenomenon called "a flat-maximum effect" can been found: with the increasement of the parameters, accuracy increases at first; but it has no significant increase from a certain threshold on, as shown in Figs 6–8.

Based on the experiment results, the evaporation rate $\rho$ and the number of ants are respectively set to 0.85 and 400 (indicated with the white dot in Fig 6), since the choice provides the best performance for the proposal in the most cases. Therefore, the choice is employed as the parameter settings for our experiments.

**Evaluation measures.**   In order to properly assess the performance of the proposed method (ELM-rules), it is tested by using the well-known 5-fold cross-validation methods. Taking into account that the predictive accuracy $Acc$ has some kind of occasionality, the test results should be assessed in statistical basis [23]. Therefore, measures like precision $Prec$ and the area under the receiver operating characteristic (ROC) curve $AUC$ are also taken into account for the performance assessment of the proposed method. If a classifier model is perfect, $AUC$ will be 1. If the model is just a random guess model, $AUC$ will be 0.5. The value of model of $AUC$ is greater, the model is more excellent. Considering the above three classification performance indicators, a composite indicator $\eta$ is used to comprehensively evaluate the TSA

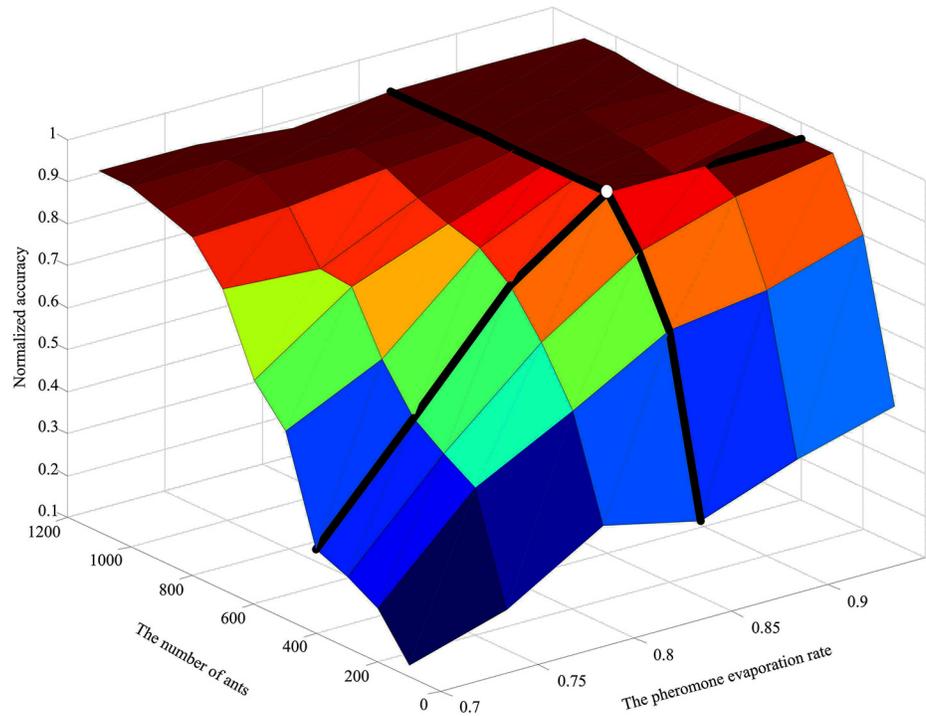

**Fig 6. Influence of the two parameters on accuracy.**

doi:10.1371/journal.pone.0130814.g006

classifier models' performance [23]. $\eta$ is defined as

$$\eta = \frac{Acc + Prec + AUC}{3}. \tag{14}$$

## Results and Discussion

**(1) Test results.** The presented method is tested by using cross-validation methods, and comparative tests are also carried out by using other relative state-of-the-art methods, including RuleFit (a rule-based ensembles method) [27], Rotation forest [28], ELM [16] and MLP. The test results are reported in Table 3 with the ROC curve shown in Fig 9.

In Table 3, *Acc* is the average validation accuracy of the 5-fold cross validation, #R is the number of the discovered rules, #T/R is the average number of rule conditions (terms) per rule, and the numbers right after the "±" symbol are the standard deviations of the corresponding evaluation measures. The parameters of the predictive models are set as follows: the parameters of RuleFit and Rotation forest are respectively set according to the reference [27] and [28]; for the ELM classifier, the hidden layer node number is set to 50; the MLP is designed with the hidden neuron number 15, and the back-propagation algorithm with the learning rate and momentum factor 0.8 and 0.7 respectively is employed.

As is shown in Table 3, the assessment results of different predictive models are different from each other, which can be summarized according to classification ability and rule list simplicity. Concerning classification ability, the classification ability of ELM is highest with the composite indicator $\eta$ getting the maximum value 0.9697; the one of MLP is lowest with $\eta$ getting the minimum value 0.9386; and the ones of RuleFit, ELM-rules and Rotation forest are respectively 0.9670, 0.9643 and 0.9643. Concerning rule list simplicity, ELM and MLP are the black-box incomprehensible predictive models; among rule-based methods, the presented method is superior to the others, and Rotation forest is the worst one.

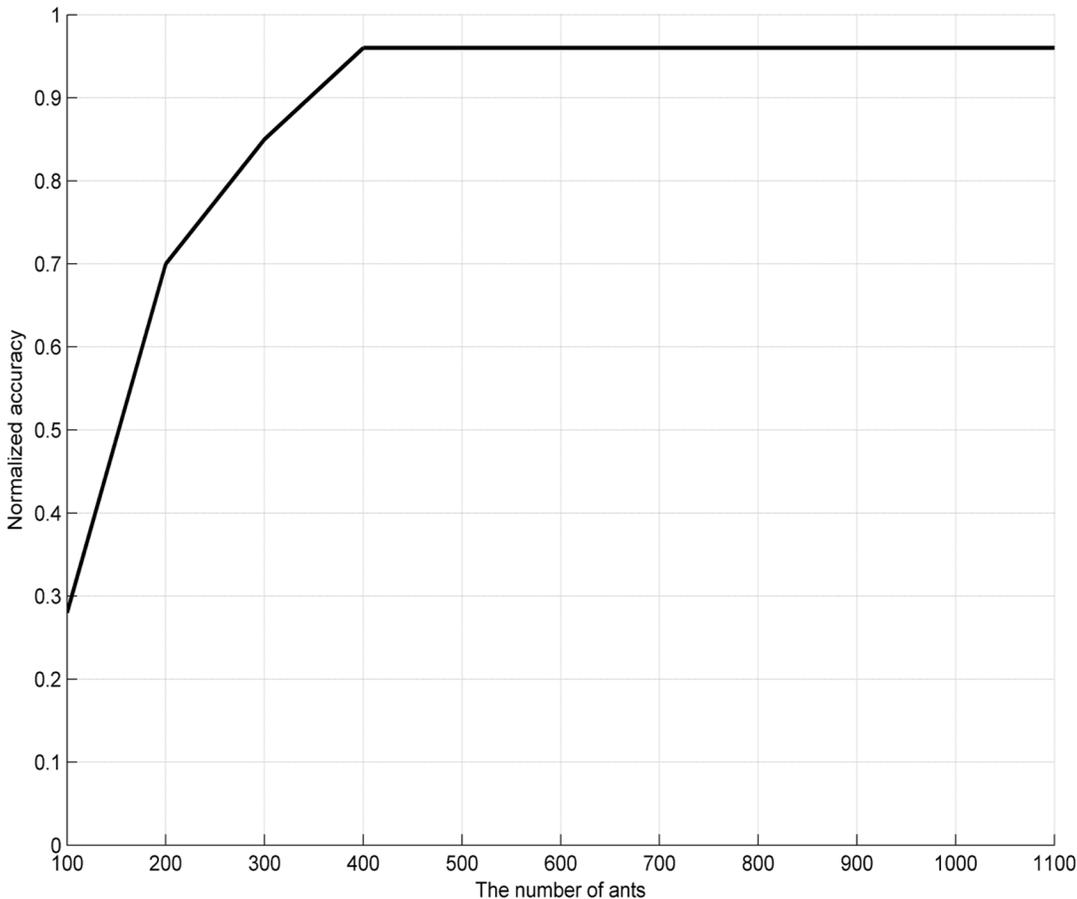

**Fig 7. Influence of the number of ants on accuracy (with $\rho$ set to 0.85).**

From Table 3, a good tradeoff between classification ability and rule list simplicity is clearly present: 1) the most accurate results are obtained by incomprehensible nonlinear ELM models; 2) the most accurate rule-based classifier, RuleFit, is slightly better than the second best rule-based classifier, ELM-rules, however, ELM-rules discovered rule lists simpler than that discovered by RuleFit since which has fewer the average number of rules and terms per rule; 3) ELM-rules and Rotation forest are roughly equivalent in terms of predictive accuracy ($\eta$ of them are equivalent to each other), and ELM-rules discovered rule lists much simpler than that discovered by Rotation forest. The reason for this is that the proposed approach extracts transient stability rules from an example sample set generated by the trained ELM-based TSA model by using IAM algorithm. By this means, the proposal combines the advantages of ELM and IAM, resulting in the good predictive performance of the obtained rules. Therefore, the conclusion can be drawn on the basis of the evidence that the proposed method is an effective way to extract the transient stability rules, since the simplicity of a rule list tends to be even more important than its predictive ability in TSA, as motivated earlier in this paper.

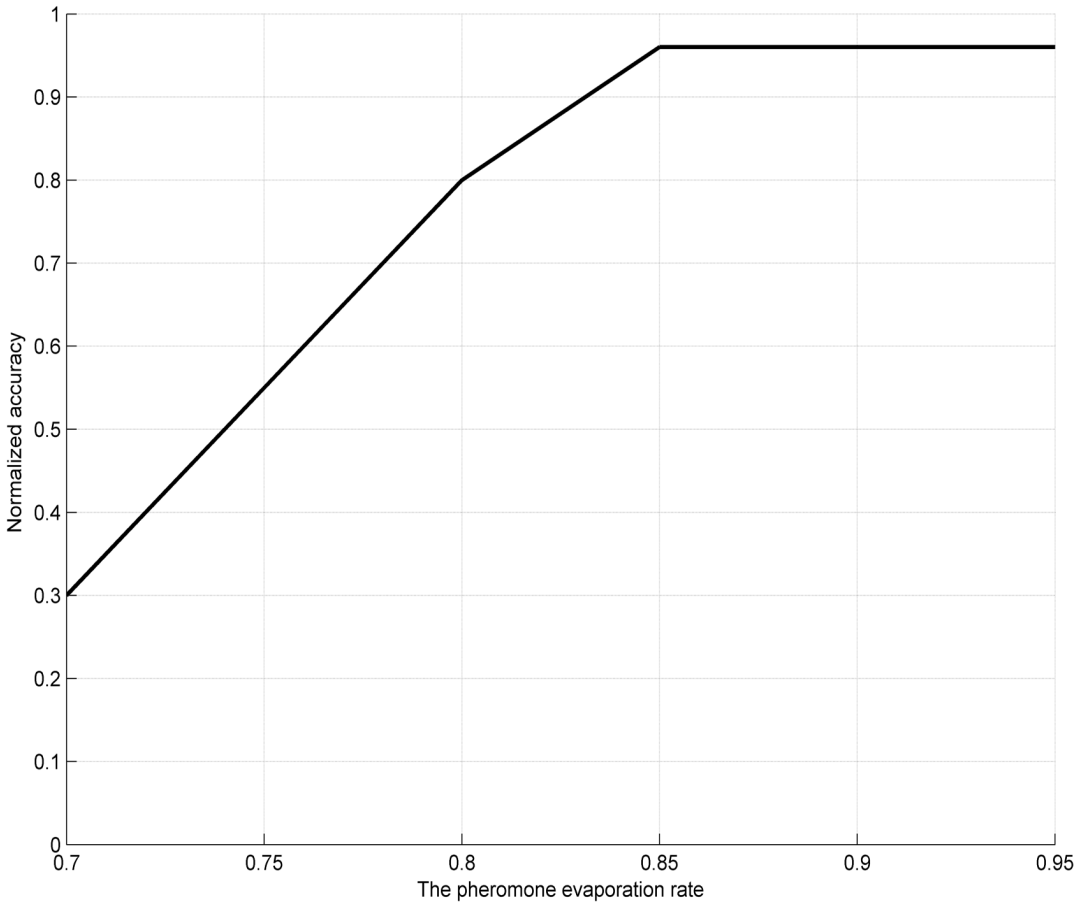

**Fig 8. Influence of the pheromone evaporation rate on accuracy (with 400 ants).**

## Case 2—The southern power system of Hebei province

In order to further justify the efficacy and applicability of the proposed approach for a more complex and practical system, it is examined on the model of a practical large power system—the southern power system of Hebei province, China. The power system covering an area of 84,000 square kilometers is a highly interconnected grid with an approximate installed capacity of 28260 MW. The modeled system comprises of 83 generators, some series compensated lines and static var compensators (SVCs).

**Table 3. Test results in Case-1.**

| Method | Acc (%) | Prec (%) | AUC | #R | #T/R | $\eta$ |
|---|---|---|---|---|---|---|
| **ELM-rules** | 97.10±0.0096 | 94.24±0.0265 | 0.9794±0.0048 | 10.5±0.24 | 3.72 | 0.9643±0.0116 |
| **RuleFit** | 97.20±0.0102 | 94.74±0.0376 | 0.9817±0.0027 | 15.2±0.41 | 4.35 | 0.9670±0.0152 |
| **Rotation forest** | 96.36±0.0105 | 0.9468±0.0226 | 0.9824±0.0097 | 18.1±0.32 | 4.81 | 0.9643±0.0102 |
| **ELM** | 97.78±0.0088 | 94.93±0.0255 | 0.9819±0.0040 | — | — | 0.9697±0.0117 |
| **MLP** | 93.24±0.0134 | 91.64±0.0317 | 0.9671±0.0068 | — | — | 0.9386±0.0145 |

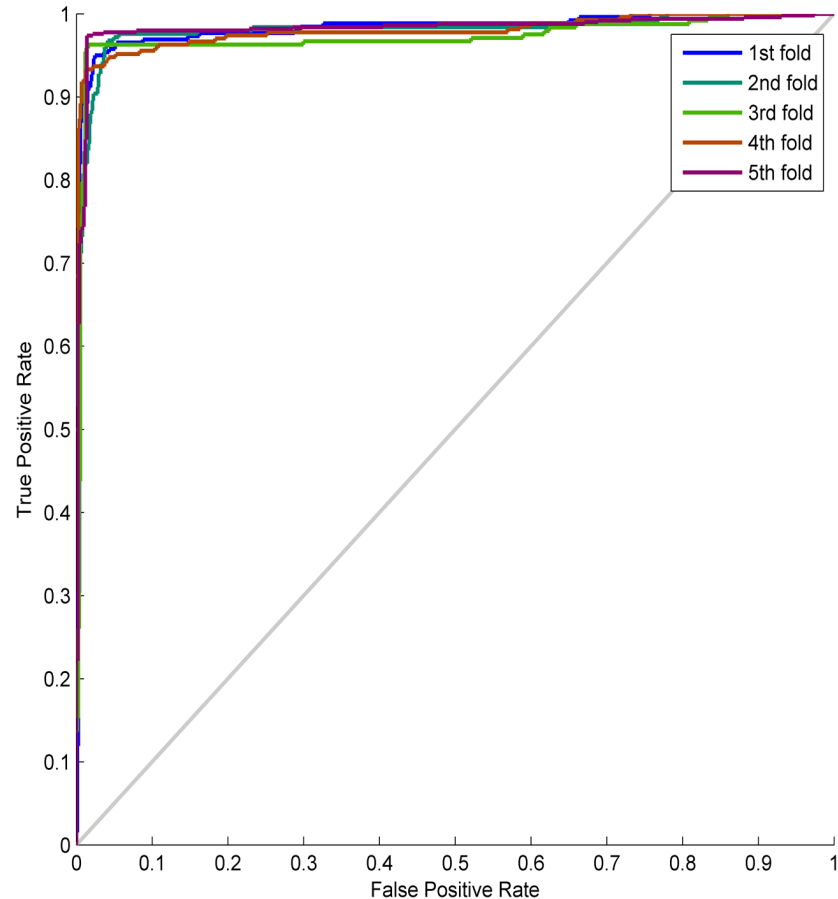

**Fig 9. The ROC curve in Case-1.**

## Generation of the sample sets

Extensive simulation has been carried out to generate the sample sets. Of all 83 generators in the system, 11 generators are modeled as the six-order model, and the excitation systems and governors are considered; others the classical machine model. The load model is represented by a comprehensive model with 40% constant-impedance and 60% constant power. In the range from 90% to 120% of the basic load level, active and reactive powers of generators are set correspondingly. Contingencies created are three-phase to ground faults, and the fault clearing times are varied from five to ten cycles. A successful reclosure of the faulted line is applied after fault clearance with no topology change from the fault. The fault locations lie at 0, 25%, 50%, and 75% of the length on transmission lines. The stability criterion is as same as in Case-1. A total of 5000 arbitrary samples are created.

## Prediction Results and Performance

In this Case, the performance of the proposed approach is also evaluated by using the 5-fold cross-validation method. By using the parameter setting method in Case-1, the evaporation rate $\rho$ and the number of ants are respectively set to 0.85 and 600 through large amounts of experiments. The ROC curve of the proposal is shown in Fig 10, and the test results of the presented method and the relative state-of-the-art methods in this Case are summarized in Table 4.

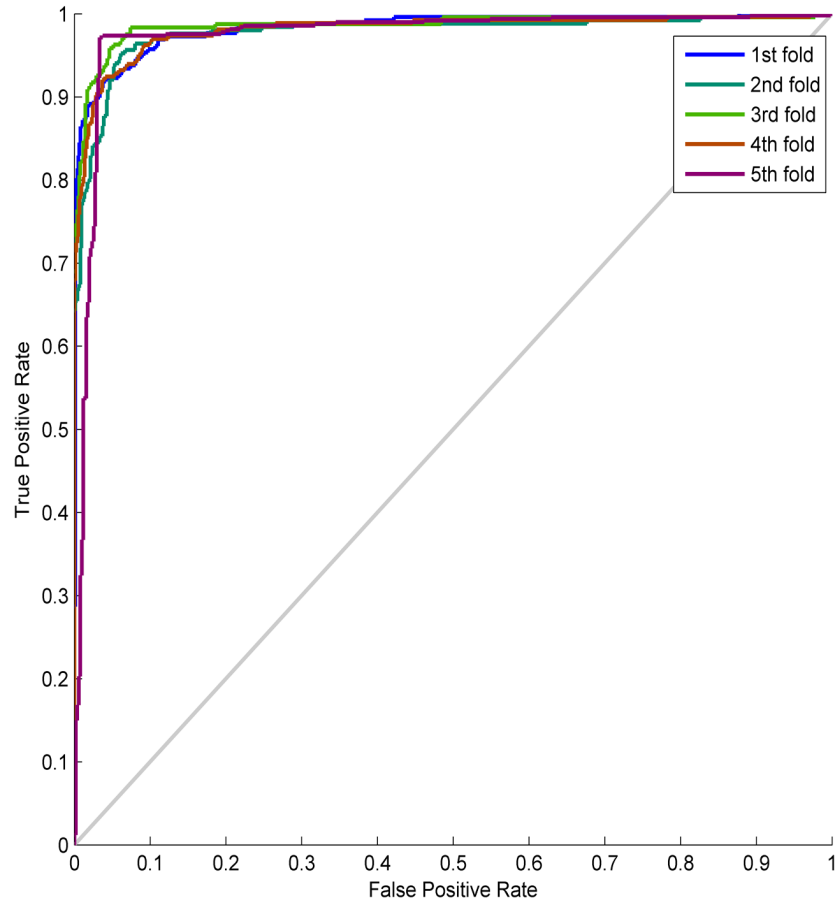

**Fig 10. The ROC curve in Case-2.**

doi:10.1371/journal.pone.0130814.g010

From Table 4, the presented results prove that the proposed algorithm is applicable to real power systems, and it is able to extract transient stability rules for a real power system—the southern power system of Hebei province. Concerning classification ability, the proposed method has roughly equivalent performance to the relative state-of-the-art works on this area; concerning rule list simplicity, the presented approach is superior to the others. Based on a comprehensive consideration of classification ability and rule list simplicity, the presented method is a good choice to solve the rule extraction problem for PRTSA.

**Table 4. Test results in Case-2.**

| Method | Acc (%) | Prec (%) | AUC | #R | #T/R | η |
|---|---|---|---|---|---|---|
| ELM-rules | 95.20±0.0065 | 92.99±0.0183 | 97.76±0.0113 | 18.6±0.18 | 6.04 | 0.9532±0.0046 |
| RuleFit | 95.32±0.0079 | 93.21±0.0231 | 97.91±0.0086 | 21.5±0.33 | 7.21 | 0.9548±0.0069 |
| Rotation forest | 95.08±0.0081 | 92.88±0.0230 | 97.93±0.0081 | 25.8±0.35 | 8.62 | 0.9530±0.0078 |
| ELM | 96.44±0.0144 | 93.42±0.0397 | 0.9782±0.0090 | — | — | 0.9589±0.0191 |
| MLP | 92.36±0.0157 | 89.13±0.0401 | 0.9620±0.0127 | — | — | 0.9256±0.0174 |

doi:10.1371/journal.pone.0130814.t004

## Conclusions

In order to improve the understandability of the PRTSA approaches, a novel rule extraction method based on ELM and IAM algorithm is presented in this paper. The key point of the proposal is that transient stability rules are extracted from an example sample set generated by the trained ELM-based TSA model by using IAM algorithm. Based on the well-known cross-validation methods, the effectiveness of the proposed method is tested on the New England 39-bus power system and the southern power system of Hebei province, and the following conclusions can be safely drawn from this work:

1. The proposed method can extract the transient stability rules not only for the New England 39-bus power system, but also for real power systems.

2. The classification ability of the proposed method is roughly equivalent to that of the relative state-of-the-art works on this area, including RuleFit and Rotation forest; however, the rule list simplicity of the proposed method is better than that of the other ones.

3. The proposed approach may find potential applications in real-time transient stability prediction of power systems. Furthermore, the methodology of rule extraction may be applied to any similar pattern classification problem in engineering field.

## Author Contributions

Conceived and designed the experiments: YL. Performed the experiments: YL GL ZW. Analyzed the data: YL GL. Contributed reagents/materials/analysis tools: YL ZW. Wrote the paper: YL.